\documentclass{article}

\usepackage{amsmath, amssymb, latexsym, amsfonts}
\usepackage{epsf}

\def\sect#1{\section{#1}\setcounter{equation}{0}}
\begin{document}

\centerline{} \vskip0.5cm \centerline{\LARGE \bf Total Variation
and  Variational Symplectic-Energy- } \vskip3pt \centerline{\LARGE \bf
Momentum integrators } \vskip0.7cm \centerline{\large
Jing-Bo Chen,$\quad$  Han-Ying Guo$\quad$   and$\quad$  Ke Wu}
\vskip3pt \centerline{\small Institute of Theoretical Physics,
Chinese Academy of Sciences} \vskip3pt \centerline{\small P.O. Box
2735, Beijing 100080, P.R. China} \vskip3pt \centerline{\small
\texttt{<chenjb><hyguo><wuke>@itp.ac.cn}} \vskip3pt


\begin{abstract}
A   discrete total variation calculus with variable time steps is
presented in this letter. Using this discrete variation calculus,
we generalize
Lee's discrete mechanics and derive variational
symplectic-energy-momentum integrators by Kane, Marsden and Ortiz.
The relationship among discrete total variation, Lee's discrete
mechanics and Kane-Marsden-Ortiz's integrators is explored.

\vskip8pt {\bf Keywords.} Total variation, Discrete mechanics,
Symplectic-energy-momentum integrators
\end{abstract}



\sect{Introduction}

In 1980's, Lee  proposed an energy-preserving discrete
mechanics with variable time steps by taking (discrete) time as
dynamical variable [\ref{td1}, \ref{td2}, \ref{td3}]. On the other
hand, motivated by the symplectic property of Lagrangian
mechanics, a version of discrete Lagrangian
mechanics has been devoloped and  variational integrators that preserve
discrete symplectic two form  have been obtained 
[\ref{m1}, \ref{m3}, \ref{v1}, \ref{v2}, \ref{w1}].
However,  variational integrators
obtained in this way have fixed time steps and consequently they
are not energy-preserving in general.

 Obviously,
the energy-preserving discrete mechanics and variational
integrator are more preferable since  solutions of the
Euler-Lagrange equations of conservative continuous systems are
not only symplectic but also energy-preserving. In order to attain
this goal, some discrete mechanics with discrete energy
conservation and  symplectic variational integrators are needed to
study.  Recently, Kane, Marsden and Ortiz have employed
appropriate  time steps to conserve a defined discrete energy and
developed what they called symplectic-energy-momentum preserving
variational integrators in [\ref{k1}]. Although their approach is
more or less related to Lee's discrete mechanics, but the discrete
energy preserving condition is not derived by the variational
principle.

The purpose of this letter is to generalize or improve these
approaches as well as to explore the relation among  discrete
total
variation, Lee's discrete mechanics and Kane-Marsden-Ortiz's
integrators. We will present a   discrete total variation calculus
with variable time steps and a discrete mechanics that is
discretely symplectic, energy preserving and has the correct
continuous limit. In fact, this discrete variation calculus and
mechanics are a generalization of Lee's discrete mechanics in
symplectic-preserving sense and can directly derive the
variational symplectic-energy-momentum integrators by Kane,
Marsden and Ortiz.

This letter is organized as follows. In the next section, we
remind total variation calculus for continuous mechanics.
 In section 3, we present a   discrete total variation calculus with variable
time steps and a discrete mechanics,  derive Kane-Marsden-Ortiz's
integrators and explore the relation among our approach, Lee's
discrete mechanics and Kane-Marsden-Ortiz's approach.
  We finish with some conclusions and remarks in Section 4.

Before ending this section, we recall very briefly the ordinary
variational principle in Lagrangian mechanics for later use.
 Suppose $Q$ denotes the extended
configuration space with coordinates $(t, q^i)$ and $Q^{(1)}$ the
first prolongation of $Q$ with coordinates $(t, q^i, \dot{q}^{i})$
[\ref{o1}]. Here $t$ denotes time and $q^i, i=1, 2, \cdots, n$
denote the position.
  Consider a Lagrangian
$L:\,\, Q^{(1)} \to \mathbb{R}$. The corresponding action functional is defined by
\begin{align}
    S(q^i(t))=\int_{a}^{b}L(t, q^i(t), \dot{q}^{i}(t))\,dt, \label{1.1}
\end{align}
where $q^{i}(t)$ is a $C^{2}$ curve in $Q$.

Hamilton's principle seeks a curve $q^i(t)$ denoted by ${\cal C}_a^b$ with
endpoints $a$ and $b$,  for which the action functional $S$ is
stationary under variations of $q^i(t)$ with fixed endpoints.
\noindent Let
\begin{align}
   V=\phi^i(t,q)\frac{\partial}{\partial q^i} \label{1.2}
\end{align}
be a vertical vector field on $Q$. Here
$q=(q^{1},\cdots, q^{n})$.
By a vertical vector field we mean a vector field on $Q$ not
involving terms of form $\xi(t,q) \frac{\partial}{\partial t}$.
Namely, time $t$ does not undergo
variation.

Let $F^{\epsilon}$ be the flow of $V$, i.e., a one-parameter group of transformations on
$Q$: $F^{\epsilon}(t,q^i)=(\tilde
{t}, \tilde{q}^i)$.
\begin{align}
 &\tilde{t}=t,\label{1.3a}\\&
\tilde{q}^i=g^i(\epsilon, t, q),\label{1.3}
\end{align}
where
\begin{align}
   \left.\frac{d}{d\epsilon}\right|_{\epsilon=0}g^i(\epsilon, t, q)
   =\phi^i(t,q):=\delta q^i(t). \label{1.4}
\end{align}
\noindent

In other words, the deformation (\ref{1.3a}-\ref{1.3}) transforms the curve
$q^i(t)$ into a family of curves $\tilde{q}^{i}(\epsilon, \tilde{t})$ in $Q$
denoted by ${{\cal
C}_\epsilon}_a^b$ 
which are determined by
\begin{align}
 &\tilde{t}=t,\label{1.5a} \\
&\tilde{q}^i=g^i(\epsilon, t, q(t)).  \label{1.5}
\end{align}
Thus, we obtain a 
(sufficiently small) set of curves ${{\cal
C}_\epsilon}_a^b$ around
${{\cal C}}_a^b$.  
Corresponding to this set of curves
there are a set of Lagrangian and action functionals
\begin{align}
S(q^i(t)) \rightarrow S(\tilde{q}^i(\epsilon, \tilde{t}))=
\int_{{a}}^{{b}}
          L( \tilde{q}^i(\epsilon, \tilde{t}), \frac{d}{d\tilde{t}}
          \tilde{q}^i(\epsilon, \tilde{t}))\,d\tilde{t}.
\label{1.6}
\end{align}
Now, we can calculate the variation of $S$ at $q(t)$ as follows
\begin{align}
\delta S=&\left.\frac{d}{d\epsilon}\right|_{\epsilon=0}S(\tilde{q}^i(\epsilon, \tilde{t}))
           \notag\\
        =&\int_{a}^{b}\left[\left(\frac{\partial L}{\partial q^i}-\frac{d}{dt}
          \frac{\partial L}{\partial \dot{q}^i}\right)\phi^i\right]dt
          +\left.\frac{\partial L}{\partial \dot{q}^i}\phi^i\right|_{a}^{b}.\label{1.7}
\end{align}
For the fixed endpoints, $\phi^i(a, q(a))=\phi^i(b, q(b))=0$, the
requirement of Hamilton's principle, $\delta S =0$, yields the
Euler-Lagrangian equation for $q(t)$
\begin{align}
 \frac{\partial L}{\partial q^i}-\frac{d}{dt}
          \frac{\partial L}{\partial \dot{q}^i}=0. \label{1.8}
\end{align}

If we drop the requirement of $\phi^i(a, q(a))=\phi^i(b, q(b))=0$,  we
can naturally obtain the Lagrangian one form on $Q^{(1)}$ from the
second term in (\ref{1.7}):
\begin{align}
\theta_{L}=\frac{\partial L}{\partial \dot{q}^i}dq^i,\label{1.9}
\end{align}
where $dq^i$ are dual to $\frac{\partial}{\partial q^j}$,
$dq^i(\frac{\partial}{\partial q^j})=\delta_j^i$.
Furthermore, 
it can be proved that the solution of (\ref{1.8}) preserves the
Lagrangian two form
\begin{align}
\omega_L:=d\theta_{L}. \label{1.10}
\end{align}%

 On the other hand,
introducing the Euler-Lagrange one form
\begin{align}
 E(q^i, \dot q^i)=\{\frac{\partial L}{\partial q^i}-\frac{d}{dt}
          \frac{\partial L}{\partial \dot{q}^{i}}\}dq^i, \label{1.11}
\end{align}
the nilpotency of $d$ leads to%
\begin{align}
dE(q^i, \dot q^i)+\frac{d}{dt}\omega_L=0. \label{1.12}\end{align}
 Namely, the
necessary and sufficient condition for symplectic structure
preserving is that the Euler-Lagrange one form is closed
[\ref{glw1}, \ref{glw2}, \ref{glw3}].



\sect{Total variation  for Lagrangian mechanics}

Consider a general  vector field on $Q$
\begin{align}
 V=\xi(t,q)\frac{\partial}{\partial t}
    +\phi^i(t,q)\frac{\partial}{\partial q^i},  \label{2.1}
\end{align}
Here $q=(q^{1},\cdots, q^{n}).$  Let 
 $F^{\epsilon}$ be the flow of $V$. The
variations of $(t,q^i)\in Q$ are described in such a way
\begin{align}
(t, q^i) \rightarrow F^{\epsilon}(t, q^i)=(\tilde{t}, \tilde{q}^i),
\label{2.2}
\end{align}
where
\begin{align}
  \tilde{t}=f(\epsilon, t, q),\qquad 
  \tilde{q}^i=g^i(\epsilon, t, q),  \label{2.3}
\end{align}
with
\begin{align}
   \left.\frac{d}{d\epsilon}\right|_{\epsilon=0}f(\epsilon, t, q)=\xi(t,q):=\delta t,
   \quad \quad \left.\frac{d}{d\epsilon}\right|_{\epsilon=0}g^i(\epsilon, t, q)
   =\phi^i(t,q):=\delta q^i.
   \label{2.4}
\end{align}
 The deformations (\ref{2.3}) transform a curve
$q^i(t)$ in $Q$ denoted by ${{\cal C}}_a^b$ into a
set of curves $\tilde{q}^i(\epsilon, \tilde{t})$ in $Q$ denoted by ${{\cal
C}_\epsilon}_{\tilde a}^{\tilde b}$,
determined by
\begin{align}
\tilde{t}=f(\epsilon, t, q(t)),\qquad 
  \tilde{q^i}=g^i(\epsilon, t, q(t)).  \label{2.5}
\end{align}
Before calculating the total variation of $S$, we introduce  the
first order prolongation of $V$ denoted as $\text{pr}^{1}V$
\begin{align}
 \text{pr}^{1}V=\xi(t,q)\frac{\partial}{\partial t}+
                \phi^i(t,q)\frac{\partial}{\partial q^i}
      +\alpha^i(t, q, \dot{q})\frac{\partial}{\partial \dot{q}^i},  \label{2.8}
\end{align}
where 
\begin{align}
 \alpha^i(t, q, \dot{q})=D_{t}\phi^i(t, q)-\dot{q}^iD_{t}\xi(t,q), \label{2.9}
\end{align}
where $D_{t}$ denotes the total derivative with respect to $t$. For example
\begin{align*}
  D_{t}\phi^k(t, q^i)=\phi_{t}^k+\phi_{q^i}^k\dot{q}^i, \quad
  \phi_{t}^k=\frac{\partial \phi^k}{\partial t}.
\end{align*}
For prolongation of vector field and the related formulae, we
refer
the reader to 
[\ref{o1}].

\noindent
Now we calculate the total variation of $S$ straightforwardly:
\begin{align}
\delta
S=&\left.\frac{d}{d\epsilon}\right|_{\epsilon=0}S(\tilde{q}^i(\epsilon,
\tilde{t}))
           \notag\\
        =&\left.\frac{d}{d\epsilon}\right|_{\epsilon=0}\int_{\tilde{a}}^{\tilde{b}}
          L(\tilde{t}, \tilde{q}^i(\epsilon, \tilde{t}), \frac{d}{d\tilde{t}}
          \tilde{q}^i(\epsilon, \tilde{t}))\,d\tilde{t}\notag\\
       =&\left.\frac{d}{d\epsilon}\right|_{\epsilon=0}\int_{a}^{b}
          L(\tilde{t}, \tilde{q}^i(\epsilon, \tilde{t}),\frac{d}{d\tilde{t}}
          \tilde{q}^i(\epsilon, \tilde{t}))
          \frac{d\tilde{t}}{dt}\,dt
          \quad\quad\quad (\tilde{t}=f(\epsilon, t, q(t)))\notag\\
        =&\int_{a}^{b}\left.\frac{d}{d\epsilon}\right|_{\epsilon=0}
           L(\tilde{t}, \tilde{q}^i(\epsilon, \tilde{t}), \frac{d}{d\tilde{t}}
          \tilde{q}^i(\epsilon, \tilde{t}))dt+\int_{a}^{b}L(t, q^i(t), \dot{q}^i(t))
           D_{t}\xi dt \notag\\
        =&\int_{a}^{b}\left[\frac{\partial L}{\partial t}\xi+
            \frac{\partial L}{\partial q^i}\phi^i+\frac{\partial L}{\partial \dot{q}^i}
            (D_{t}\phi^i-\dot{q}^iD_{t}\xi)\right]\,dt
            +\int_{a}^{b}L\,D_{t}\xi\,dt \notag\\
         \begin{split}
         =&\int_{a}^{b}\left[\left(\frac{\partial L}{\partial t}+\frac{d}{dt}\left(
           \frac{\partial L}{\partial \dot{q}^i}\dot{q}^i-L\right)\right)\xi+\left(
          \frac{\partial L}{\partial q^i}-\frac{d}{dt}\frac{\partial L}{\partial \dot{q}^i}
          \right)\phi^i\right]\,dt \\
          &\quad +\left.\left[\left(L-\frac{\partial L}{\partial \dot{q}^i}
           \dot{q}^i\right)\xi
           +\frac{\partial L}{\partial \dot{q}^i}\phi^i\right]\right|_{a}^{b}.
         \end{split} \label{2.10}
\end{align}
Here  we have made use of  (\ref{2.4}), (\ref{2.8}), (\ref{2.9}) and %
$$\left.\frac{d}{d\epsilon}\right|_{\epsilon=0}
\frac{d\tilde{t}}{dt}=\frac{d}{dt}\left.\frac{d}{d\epsilon}\right|_{\epsilon=0}
\tilde{t} =D_{t}\xi.$$
\noindent If $\xi(a, q(a))=\xi(b, q(b))=0$ and $\phi^i(a,
q(a))=\phi^i(b, q(b))=0$,
 the requirement of $\delta S
=0$ yields the equation from $\xi$, the variation along
the base manifold, i.e. the time,
\begin{align}
\frac{\partial L}{\partial t}+\frac{d}{dt}\left(
           \frac{\partial L}{\partial \dot{q}^i}\dot{q}^i-L\right)=0,
\label{2.12}
\end{align}
 and
the Euler-Lagrange equation from $\phi^i$, the 
 variation along the fibre, i.e. the configuration space,
\begin{align}
 \frac{\partial L}{\partial q^i}-\frac{d}{dt}
          \frac{\partial L}{\partial \dot{q}^i}=0. \label{2.13}
\end{align}
\noindent Here $\xi$ and $\phi^i$  are 
regarded as independent components of
the total variation. However, there is another decomposition for
the independent components, i.e. the 
vertical and
horizontal variations, see {\em Remark} 2 below.

If $L$ does not depend on $t$ explicitly, i.e.,  $L$ is conservative, 
  $\frac{\partial L}{\partial t}=0$, then
(\ref{2.12}) becomes the energy conservation law
\begin{align}
\frac{d}{dt}H=0, \qquad H:=\left(
           \frac{\partial L}{\partial \dot{q}^i}\dot{q}^i-L\right).
\label{2.14}
\end{align}
 We expand the left-hand side of (\ref{2.14}) and obtain
\begin{align}
\frac{d}{dt}\left(
           \frac{\partial L}{\partial \dot{q}^i}\dot{q}^i-L\right)=
-\left(\frac{\partial L}{\partial q^i}-\frac{d}{dt}
          \frac{\partial L}{\partial \dot{q}^i}\right)\dot{q}^i.
\label{2.15}
\end{align}
Thus, for a conservative $L$, energy conservation is a consequence
of Euler-Lagrange equation. This agrees with Noether theorem,
which states that the  characteristic of an infinitesimal symmetry
of  the action functional $S$ is that of a conservation law for
the Euler-Lagrange equation. For a conservative $L$,
$\frac{\partial}{\partial t}$ is an infinitesimal symmetry of  the
action functional $S$, and its characteristic is $-\dot{q}^i$.
From Noether theorem, there exits a corresponding conservation law
in the characteristic form
\begin{align}
 -\left(\frac{\partial L}{\partial q^i}-\frac{d}{dt}
\frac{\partial L}{\partial \dot{q}^i}\right)\dot{q}^i=0.
\label{2.16}
\end{align}
If we drop the requirement
\begin{align}
 \xi(a, q(a))=\xi(b, q(b))=0, \qquad
 \phi^i(a, q(a))=\phi^i(b, q(b))=0,
\end{align}
we can define the extended Lagrangian one form on $Q^{(1)}$ from
the second term in (\ref{2.10})
\begin{align}
\vartheta_{L}:=\left(L-\frac{\partial L}{\partial \dot{q}^i}
           \dot{q}^i\right)dt +\frac{\partial L}{\partial \dot{q}^i}dq^i \label{2.17}.
\end{align}
Suppose $g^i(t, v_{q^i})$ is a solution of (\ref{2.13}) depending
on the initial condition $v_{q^i}\in Q^{(1)}$. Restricting
$\tilde{q}^i(\epsilon, \tilde{t})$ to the solution space of
(\ref{2.13}) and using the same method in [\ref{m1}], it can be
proved that the extended symplectic two form is preserved
\begin{align}
(\text{pr}^{1}g^i)^{*}\Omega_{L}=\Omega_{L},\qquad
\Omega_L:=d\vartheta_L. \label{2.22}
\end{align}
where $\text{pr}^{1}g^i(s, v_{q^i}))=(s,g^i(s, v_{q^i}),
\frac{d}{ds}g^i(s, v_{q^i}))$ denotes the first order prolongation
of $g^i(s, v_{q^i})$ [\ref{o1}].

 \vskip10pt 
 \noindent
 {\em Remark} \,1. \, If $\xi$ in
(\ref{2.1}) is independent of $q$, the deformations in
(\ref{2.3}) are called fiber-preserving. In this case, the domain
of definition of $\tilde{q}^i(\epsilon, \tilde{t})$ only depends
on the deformations (\ref{2.3}). While in the general case, the
domain of definition of $\tilde{q}^i(\epsilon, \tilde{t})$ not
only depends on
the deformations (\ref{2.3}) but also on $q^i(t)$. 

\vskip10pt
\noindent
{\em Remark}\, 2. \, Using the identity 
\begin{align}
\frac{\partial L}{\partial t}+\frac{d}{dt}\left(
           \frac{\partial L}{\partial \dot{q}^i}\dot{q}^i-L\right)
=-\left(\frac{\partial L}{\partial q^i}-\frac{d}{dt}\frac{\partial
L}{\partial \dot{q}^i}
          \right)\dot{q}^i, \label{2.24a}
\end{align}
the Eq. (\ref{2.10}) becomes
\begin{align}
\delta S=\int_{a}^{b}\left(\frac{\partial L}{\partial
q^i}-\frac{d}{dt}
         \frac{\partial L}{\partial \dot{q}^i}\right)(\phi^i-\xi \dot{q}^i)dt
          +\left.\left[\frac{\partial L}{\partial \dot{q}^i}(\phi^i-\xi \dot{q}^i)
           \right]\right|_{a}^{b}+ \left.(L\xi)\right|_{a}^{b}.\label{2.24}
\end{align}

According to (\ref{2.4}), $\phi^i=\delta q^i$ should be regarded
as the total variation of $q^i$, $\delta q^i=\delta_V q^i
+\delta_H q^i$, since the variation of $t$ also induces the
variation of $q^i$ denoted as $\delta_H q^i$, the horizontal
variation of $ q^i$. Due to $\xi=\delta t$ in (\ref{2.4}), the
horizontal variation of $ q^i$ should be  $\delta_H q^i=\xi
\dot{q}^i$ and consequently $\phi^i-\xi \dot{q}^i$ is interpreted
as vertical variation $\delta_V q^i$, i.e. the variation of
$q^i(t)$
at the moment $t$ (see, for example [\ref{c1}]). 
Therefore, the first two terms in (\ref{2.24}) come from vertical
variation $\delta_V q^i$ and the last term comes from horizontal
variation $\delta t$.  The horizontal variation of $S$ with
respect to the horizontal variation  $\delta_H q^i=\xi \dot{q}^i$
just gives rise to the identity (\ref{2.24a}).


\vskip5pt
\sect{Discrete mechanics and variational integrators with variable
time steps }

 In this section, by means of a calculus of discrete total variations we develop
 a discrete Lagrangian mechanics, which includes the boundary terms in Lee's
discrete mechanics that give rise to the discrete version of
symplectic preserving. The discrete variation calculus is mainly
analog to Lee's idea that (discrete) time is regarded as dynamical
variable, i.e. the time steps are variable [\ref{td1}, \ref{td2},
\ref{td3}]. And the vertical part of this discrete variation
calculus is similar to the one in [\ref{k1}, \ref{m1}, \ref{m3},
\ref{v1}, \ref{v2}, \ref{w1}]. 
Using this calculus
for discrete total  variations
we  naturally derive
 Kane-Marsden-Ortiz's integrators.

 We use $Q \times Q$ for the discrete version
of the first prolongation for the extended configuration space
$Q$. A point $(t_{0}, q_{0}; t_{1}, q_{1})\in Q\times Q$
{\footnote{In this section $q$ is an abbreviation of
$(q^{1}, q^{2}, \cdots, q^{n})$.}}corresponds to a tangent vector
$\frac{q_{1}-q_{0}}{t_{1}-t_{0}}$. A discrete Lagrangian is
defined to be $\mathbb{L}:\,Q\times Q\to \mathbb{R}$ and the
corresponding action to be
\begin{align}
 \mathbb{S}=\sum_{k=0}^{N-1}\mathbb{L}(t_{k}, q_{k}, t_{k+1}, q_{k+1})
   (t_{k+1}-t_{k}). \label{3.1}
\end{align}
The discrete variational principle in total variation  is to
extremize $\mathbb{S}$ for variations of both $q_{k}$ and $t_{k}$
with  fixed endpoints $(t_{0}, q_{0})$ and $(t_{N}, q_{N})$. This
discrete variational principle  determines a discrete flow
$\Phi:\,Q\times Q\to Q\times Q$ by
\begin{align}
   \Phi(t_{k-1}, q_{k-1}, t_{k}, q_{k})=(t_{k}, q_{k},t_{k+1}, q_{k+1}). \label{3.2}
\end{align}
Here $(t_{k+1}, q_{k+1})$ are found from the following discrete
Euler-Lagrange equation, i.e. the variational integrator, and the
discrete energy conservation law (for conservative L)
\begin{align}
 &(t_{k+1}-t_{k})D_{2}\mathbb{L}(t_{k}, q_{k}, t_{k+1}, q_{k+1})
  +(t_{k}-t_{k-1})D_{4}\mathbb{L}(t_{k-1}, q_{k-1}, t_{k}, q_{k})=0,
 \label{3.3}
\end{align}
and
\begin{align}
 \begin{split}
 &(t_{k+1}-t_{k})D_{1}\mathbb{L}(t_{k}, q_{k}, t_{k+1}, q_{k+1})
   +D_{3}\mathbb{L}(t_{k-1}, q_{k-1}, t_{k}, q_{k})(t_{k}-t_{k-1}) \\
 &\quad -\mathbb{L}(t_{k}, q_{k}, t_{k+1}, q_{k+1})+
\mathbb{L}(t_{k-1}, q_{k-1}, t_{k}, q_{k})=0,
\end{split}\label{3.4}
\end{align}
for all $k\in\{1, 2, \cdots, N-1\}$. Here $D_{i}$ denotes the
partial derivative of $\mathbb{L}$ with respect to the $i$th
argument. The Eq. (\ref{3.3}) is the discrete Euler-Lagrange
equation. 
The Eq. (\ref{3.4}) is the discrete energy conservation law for a
conservative $\mathbb{L}$. The integrator (\ref{3.3})-(\ref{3.4})
is just Kane-Marsden-Ortiz's integrator.

Using the discrete flow $\Phi$, the Eqs. (\ref{3.3}) and (\ref{3.4}) become
respectively
\begin{align}
  &(t_{k+1}-t_{k})D_{2}\mathbb{L}\circ\Phi +
     (t_{k}-t_{k-1})D_{4}\mathbb{L}=0, \label{3.5}\\
  &((t_{k+1}-t_{k})D_{1}\mathbb{L}-\mathbb{L})\circ\Phi +
   D_{3}\mathbb{L}+\mathbb{L}=0. \label{3.6}
\end{align}
If $(t_{k+1}-t_{k})D_{2}\mathbb{L}$ and $(t_{k+1}-t_{k})D_{1}\mathbb{L}-\mathbb{L}$
are invertible, the Eqs.
(\ref{3.5}) and (\ref{3.6}) determine the discrete flow $\Phi$ under
the consistency condition
\begin{align}
   ((t_{k+1}-t_{k})D_{1}\mathbb{L}-\mathbb{L})^{-1}\circ (D_{3}\mathbb{L}+\mathbb{L})
    =((t_{k+1}-t_{k})D_{2}\mathbb{L})^{-1}\circ (t_{k}-t_{k-1})D_{4}\mathbb{L}.
\end{align}

Now we prove that the discrete flow $\Phi$ preserves a discrete
version of the extended Lagrange two form $\Omega_{L}$. 
As in continuous case,
we calculate $d\mathbb{S}$ for variations with varied endpoints.
\begin{align}
 &d\mathbb{S}(t_{0}, q_{0}, \cdots, t_{N}, q_{N})\cdot (\delta t_{0}, \delta q_{0},
      \cdots, \delta t_{N}, \delta q_{N})\notag \\
  &=\sum_{k=0}^{N-1}(D_{2}L(t_{k}, q_{k}, t_{k+1}, q_{k+1})\delta q_{k}+
        D_{4}L(t_{k}, q_{k}, t_{k+1}, q_{k+1})\delta q_{k+1})(t_{k+1}-t_{k})\notag \\
     &\quad + \sum_{k=0}^{N-1}(D_{1}L(t_{k}, q_{k}, t_{k+1}, q_{k+1})\delta t_{k}+
        D_{3}L(t_{k}, q_{k}, t_{k+1}, q_{k+1})\delta t_{k+1})(t_{k+1}-t_{k})\notag \\
     &\quad + \sum_{k=0}^{N-1}L(t_{k}, q_{k}, t_{k+1}, q_{k+1})(\delta t_{k+1}
       -\delta t_{k})\notag\\
  &=\sum_{k=0}^{N-1}D_{2}L(t_{k}, q_{k}, t_{k+1}, q_{k+1})(t_{k+1}-t_{k})\delta q_{k}
     \notag\\
     &\quad +\sum_{k=1}^{N} D_{4}L(t_{k-1}, q_{k-1}, t_{k}, q_{k})
          (t_{k}-t_{k-1})\delta q_{k}\notag \\
     &\quad +\sum_{k=0}^{N-1}D_{1}L(t_{k}, q_{k}, t_{k+1}, q_{k+1})(t_{k+1}-t_{k})
        \delta t_{k}\notag \\
     &\quad +\sum_{k=1}^{N} D_{3}L(t_{k-1}, q_{k-1}, t_{k}, q_{k})(t_{k}-t_{k-1})
        \delta t_{k}\notag \\
     &\quad + \sum_{k=0}^{N-1}L(t_{k}, q_{k}, t_{k+1}, q_{k+1})(-\delta t_{k})
       +\sum_{k=1}^{N}L(t_{k-1}, q_{k-1}, t_{k}, q_{k})\delta t_{k}\notag
       \end{align}
       \begin{align}
  &=\sum_{k=1}^{N-1}(D_{2}L(t_{k}, q_{k}, t_{k+1}, q_{k+1})(t_{k+1}-t_{k})
                     +D_{4}L(t_{k-1}, q_{k-1}, t_{k}, q_{k})
                     (t_{k}-t_{k-1}))\delta q_{k}\notag\\
    &\quad +\sum_{k=1}^{N-1}(D_{1}L(t_{k}, q_{k}, t_{k+1}, q_{k+1})(t_{k+1}-t_{k})
                     +D_{3}L(t_{k-1}, q_{k-1}, t_{k}, q_{k})
                     (t_{k}-t_{k-1})\notag\\[-10pt]
       &\quad \quad\quad \quad \,\,+L(t_{k-1}, q_{k-1}, t_{k}, q_{k})
      -L(t_{k}, q_{k}, t_{k+1}, q_{k+1}))\delta t_{k}\notag\\
    &\quad +D_{2}L(t_{0}, q_{0}, t_{1}, q_{1})(t_{1}-t_{0})\delta q_{0}
           +D_{4}L(t_{N-1}, q_{N-1}, t_{N}, q_{N})(t_{N}-t_{N-1})\delta q_{N}\notag\\
    &\quad +(D_{1}L(t_{0}, q_{0}, t_{1}, q_{1})(t_{1}-t_{0})
               -L(t_{0}, q_{0}, t_{1}, q_{1}))\delta t_{0}\notag\\
     &\quad  +(D_{3}L(t_{N-1}, q_{N-1}, t_{N}, q_{N})(t_{N}-t_{N-1})
               +L(t_{N-1}, q_{N-1}, t_{N}, q_{N}))\delta t_{N}. \label{3.8}
\end{align}
We can see that the last four terms in (\ref{3.8}) come from the boundary variations.
Based on the
boundary variations, we can define two one forms on $Q\times Q$
\begin{align}
\begin{split}
&\theta_{\mathbb{L}}^{-}(t_{k}, q_{k}, t_{k+1}, q_{k+1})\\
                   &\, =(D_{1}\mathbb{L}(t_{k}, q_{k}, t_{k+1}, q_{k+1})
            (t_{k+1}-t_{k})-\mathbb{L}(t_{k}, q_{k}, t_{k+1}, q_{k+1}))dt_{k}\\
          &\, \quad +D_{2}\mathbb{L}(t_{k}, q_{k}, t_{k+1}, q_{k+1})
                (t_{k+1}-t_{k}) dq_{k}, \label{3.9}
\end{split}
\end{align}
and
\begin{align}
\begin{split}
 &\theta_{\mathbb{L}}^{+}(t_{k}, q_{k}, t_{k+1}, q_{k+1})\\
             &\,    =(D_{3}\mathbb{L}(t_{k}, q_{k}, t_{k+1}, q_{k+1})
            (t_{k+1}-t_{k})+\mathbb{L}(t_{k}, q_{k}, t_{k+1}, q_{k+1}))dt_{k+1}\\
          &\, \quad +D_{4}\mathbb{L}(t_{k}, q_{k}, t_{k+1}, q_{k+1})
                (t_{k+1}-t_{k}) dq_{k+1}. \label{3.10}
\end{split}
\end{align}
Here we have employed the notations in [\ref{m1}]. We regard the
pair
$(\theta_{\mathbb{L}}^{-}, \theta_{\mathbb{L}}^{+})$ as 
the discrete version of the extended Lagrange one form
$\vartheta_{L}$ defined in (\ref{2.17}).

Now we parameterize the solutions of the discrete variational
principle by the initial condition $(t_{0}, q_{0}, t_{1}, q_{1})$
and  restrict $\mathbb{S}$ to that solution space.  Then Eq.
(\ref{3.8}) becomes
\begin{align}
\begin{split}
 d&\mathbb{S}(t_{0}, q_{0}, \cdots, t_{N}, q_{N})\cdot (\delta t_{0}, \delta q_{0},
      \cdots, \delta t_{N}, \delta q_{N})\\
 &=\theta_{\mathbb{L}}^{-}(t_{0}, q_{0}, t_{1}, q_{1})\cdot
                        (\delta t_{0}, \delta q_{0},  \delta t_{1}, \delta q_{1})\\
   &\quad +\theta_{\mathbb{L}}^{+}(t_{N-1}, q_{N-1}, t_{N}, q_{N})\cdot
                (\delta t_{N-1}, \delta q_{N},  \delta t_{N-1}, \delta q_{N-1})\\
 &=\theta_{\mathbb{L}}^{-}(t_{0}, q_{0}, t_{1}, q_{1})\cdot
                        (\delta t_{0}, \delta q_{0},  \delta t_{1}, \delta q_{1})\\
   &\quad +(\Phi^{N-1})^{*}\theta_{\mathbb{L}}^{+}(t_{0}, q_{0}, t_{1}, q_{1})\cdot
        (\delta t_{0}, \delta q_{0},  \delta t_{1}, \delta q_{1}). \label{3.11}
\end{split}
\end{align}
From (\ref{3.11}), we can obtain
\begin{align}
d\mathbb{S}=\theta_{\mathbb{L}}^{-}+(\Phi^{N-1})^{*}\theta_{\mathbb{L}}^{+}.
\label{3.12}
\end{align}
The Eq. (\ref{3.12}) holds for arbitrary $N>1$. By taking N=2, it
leads to
\begin{align}
d\mathbb{S}=\theta_{\mathbb{L}}^{-}+\Phi^{*}\theta_{\mathbb{L}}^{+}.
\label{3.13}
\end{align}
Taking exterior differentiation of (\ref{3.13}), it follows
that
\begin{align}
  \Phi^{*}(d\theta_{\mathbb{L}}^{+})=-d\theta_{\mathbb{L}}^{-}. \label{3.14}
\end{align}
From the definition of $\theta_{\mathbb{L}}^{-}$ and $\theta_{\mathbb{L}}^{+}$, we
know that
\begin{align}
   \theta_{\mathbb{L}}^{-}+\theta_{\mathbb{L}}^{+}=d(\mathbb{L}(t_{k+1}-t_{k})).\label{3.15}
\end{align}
Taking exterior differentiation of (\ref{3.15}), we obtain
$d\theta_{\mathbb{L}}^{+}=-d\theta_{\mathbb{L}}^{-}$.  Define
\begin{align}
 \Omega_{\mathbb{L}}\equiv d\theta_{\mathbb{L}}^{+}=
-d\theta_{\mathbb{L}}^{-}.\label{3.16}
\end{align}
Finally, we have shown that the discrete flow $\Phi$ preserves the
discrete extended Lagrange two form $\Omega_{\mathbb{L}}$.
\begin{align}
  \Phi^{*}(\Omega_{\mathbb{L}})=\Omega_{\mathbb{L}}. \label{3.17}
\end{align}

Now we show that the variational integrator (\ref{3.3}), the
discrete energy conservation law (\ref{3.4}) and the discrete
extended Lagrange two form $\Omega_{\mathbb{L}}$ converge to their
continuous counterparts as $t_{k+1}\to t_{k},\,\, t_{k-1}\to
t_{k}$.

Consider a conservative Lagrangian $L(q, \dot{q})$. For simplicity, we choose the
discrete Lagrangian as
\begin{align}
\mathbb{L}(t_{k}, q_{k}, t_{k+1}, q_{k+1})=L(q_{k}, \frac{q_{k+1}-q_{k}}
{t_{k+1}-t_{k}}).\label{3.18}
\end{align}
The variational integrator (\ref{3.3}) becomes
\begin{align}
   &\frac{\partial L}{\partial q_{k}}( q_{k}, \Delta_{t}q_{k})
   -\frac{1}{t_{k+1}-t_{k}}\left(\frac{\partial L}{\partial \Delta_{t}q_{k}}
    (q_{k}, \Delta_{t}q_{k})-\frac{\partial L}{\partial \Delta_{t}q_{k-1}}
    ( q_{k-1}, \Delta_{t}q_{k-1} )\right)=0, \label{3.19}
\end{align}
where $\Delta_{t}q_{k}=\frac{q_{k+1}-q_{k}}{t_{k+1}-t_{k}},\quad
\Delta_{t}q_{k-1}=\frac{q_{k}-q_{k-1}}{t_{k}-t_{k-1}}$.

It is easy to see that as $t_{k+1}\to
t_{k},\,\,  t_{k-1}\to t_{k}$,  the Eq. (\ref{3.19}) converges to
\begin{align}
\frac{\partial L}{\partial q_{k}}-\frac{d}{dt}
\frac{\partial L}{\partial \dot{q}_{k}}=0. \label{3.20}
\end{align}
\noindent
The discrete energy conservation law (\ref{3.4}) becomes
\begin{align}
\frac{E_{k+1}-E_{k}}{t_{k+1}-t_{k}}=0, \label{3.21}
\end{align}
where
\begin{align*}
  &E_{k+1}=\frac{\partial L}{\partial \Delta_{t}q_{k}}\Delta_{t}q_{k}
           -L(q_{k}, \frac{q_{k+1}-q_{k}}{t_{k+1}-t_{k}})\\
  &E_{k}=\frac{\partial L}{\partial \Delta_{t}q_{k-1}}\Delta_{t}q_{k-1}
           -L(q_{k-1}, \frac{q_{k}-q_{k-1}}{t_{k}-t_{k-1}}).
\end{align*}
The Eq. (\ref{3.21}) converges to
\begin{align}
\frac{d}{dt}\left(\frac{\partial L}{\partial \dot{q}_{k}}\dot{q}_{k}-L\right)=0
\label{3.22}
\end{align}
as $t_{k+1}\to t_{k},\,\, t_{k-1}\to t_{k}$.

Now we consider the discrete extended Lagrange two form
$\Omega_{\mathbb{L}}$ defined by (\ref{3.16}). Under the
discretization (\ref{3.18}), the discrete extended Lagrange one
form $\theta_{\mathbb{L}}^{+}$ defined in (\ref{3.10}) becomes
\begin{align}
\theta_{\mathbb{L}}^{+}=\left(L(q_{k}, \Delta_{t}q_{k})
-\frac{\partial L}{\partial \Delta_{t}q_{k}}
           \Delta_{t}q_{k}\right)dt_{k+1} +
        \frac{\partial L}{\partial \Delta_{t}q_{k} }dq_{k+1}. \label{3.23}
\end{align}
From (\ref{3.23}), we can deduce that $\theta_{\mathbb{L}}^{+}$
converges to the continuous Lagrangian one form $\vartheta_{L}$
defined by (\ref{2.17}) as $t_{k+1}\to t_{k}, \,\,t_{k-1}\to
t_{k}$. Thus, we obtain
\begin{align}
  \Omega_{\mathbb{L}}=d\theta_{\mathbb{L}}^{+}\to d\vartheta_{L}=\Omega_{L},\quad
 (t_{k+1}\to t_{k}, \,\,t_{k-1}\to t_{k}). \label{3.24}
\end{align}

The variational integrator (\ref{3.3}) with fixed time steps does
not conserve the discrete energy exactly in general but the
computed energy will not have secular variation [\ref{g1},
\ref{s1}]. In some cases such as in discrete mechanics proposed by
Lee in
[\ref{td1}, \ref{td2}, \ref{td3}], 
the integrator (\ref{3.3}) is required to conserve the discrete
energy (\ref{3.4}) by varying the time steps. In other words, the
time steps can be chosen according to (\ref{3.4}) so as to the
integrator (\ref{3.3}) conserves the discrete energy (\ref{3.4}).
The resulting
integrator also conserves the discrete extended Lagrange two form
$d\theta_{\mathbb{L}}^{+}$.
 This fact had not been
discussed in Lee's discrete mechanics.



\sect{Conclusions}

We have presented the calculus of total variation problem
for discrete mechanics with variable time steps referring the one
for continuous mechanics
 in this letter.
Using the calculus for discrete total  variations, we have proved
that Lee's discrete mechanics is symplectic and derived
Kane-Marsden-Ortiz's integrators. It is well acknowledged, an
energy-preserving variational integrator is a more preferable and
natural candidate of approximations for conservative
Euler-Lagrangian equation since the solution of conservative
Euler-Lagrangian equation is not only symplectic but also
energy-preserving.

As is mentioned, Kane-Marsden-Ortiz's integrators are closely
related to the discrete mechanics proposed by Lee [\ref{td1},
\ref{td2}, \ref{td3}]. 
In Lee's discrete mechanics, the difference equations are just
Kane-Marsden-Ortiz's integrators.
 However, Lee's difference equations are solved as
boundary value problems, while Kane-Marsden-Ortiz's integrators
are solved as initial value problems.

Finally, it should be mentioned that two of the authors (HYG and
KW) and their collaborators have presented a difference discrete
variational calculus and the discrete version of Euler-Lagrange
cohomology  for vertical variation problems in both Lagrangian and
Hamiltonian formalism for discrete mechanics and field theory very
recently [\ref{glw1}, \ref{glww}]. In their approach, the
difference operator with fixed step-length  is regarded as an
entire geometric object. The advantages of this approach have
already be seen in the last section in the course of taking
continuous limits although the difference operator $\Delta_t$ in
(\ref{3.18}) is of variable step-length. This approach should be
able to be generalized for the  discrete total variation problems.





\begin{thebibliography} {99}

\bibitem{c1} \label{c1}
R. Courant and D. Hilbert, Methods of Mathematical Physics, {\em Interscience, New
York, } (1953).

\bibitem{l1}\label{td1}
T.D. Lee, Can time be a discrete dynamical variable? {\em Phys.
Lett.}, {\bf 122B}, (1982).

\bibitem{l2}\label{td2}
T.D. Lee, Difference equations and conservation laws, {\em J.
Statis. Phys.}, {\bf 46}, 843-860, (1987).

\bibitem{l3}\label{td3}
T.D. Lee, Discrete mechanics, {\em Lectures given at the
International School of Subnuclear Physics, Erice, August 1983}.

\bibitem{g1}\label{g1}
Z. Ge and J.E. Marsden, Lie-Poisson integrators and Lie-Poisson Hamilton-Jacobi
theory, {\em Phys. Lett. A}, {\bf 133}, 134-139, (1988).


\bibitem{glw1}\label{glw1} H.Y. Guo, Y.Q. Li and K. Wu, On symplectic and multisymplectic
structures and their discrete versions in Lagrangian formalism,
{\em Comm. Theor. Phys.}, {\bf 35} (2001) 703-710.

\bibitem{glw2}\label{glw2} H.Y. Guo,
Y.Q. Li and K. Wu, A note on symplectic algorithms, {\em Comm.
Theor. Phys.}, {\bf 36} (2001) 11-18.

\bibitem{glw3}\label{glw3} H.Y. Guo, Y.Q. Li, K. Wu and S.K. Wang,
Symplectic, multisymplectic structures and Euler-Lagrange
cohomology, Preprint AS-ITP-2001-009, April, 2001.
arXiv:hep-th/0104140.

\bibitem{glww}\label{glww}
 H.Y. Guo, Y.Q. Li, K. Wu and S.K. Wang, Difference discrete variational principle,
 Euler-Lagrange cohomology and
 symplectic, multisymplectic structures, Preprint arXiv:hep-th/0106001.

\bibitem{k1}\label{k1}
C. Kane, J. E. Marsden and M. Ortiz,  Symplectic-energy-momentum preserving
variational integrators, {\em J. Math. Phys.}, {\bf 40}, 3353-3371.


\bibitem{m1}\label{m1}
J.E. Marsden, G.P. Patrick and S. Shkoller, Multisymplectic geometry,
variational integrators, and nonlinear PDEs, {\em Comm. Math. Phys.},
{\bf 199}, 351-395 (1998).


\bibitem{m3}\label{m3}
J. Moser and A.P. Veselov, Discrete versions of some classical integrable systems
and factorization of matrix polynomials, {\em Commun. Math. Phys.}, {\bf 139},
217-243, (1991).

\bibitem{olver}\label{o1}
 P. J. Olver, Applications of Lie groups to differential equations.
GTM 107,{\em Springer-Verlag.} (1993)

\bibitem{s1}\label{s1}
J.M. Sanz-serna and M. Calvo, Numerical Hamiltonian problems. {\em
Chapman and Hall Lodon}, (1994).

\bibitem{v1}\label{v1}
A.P. Veselov, Integrable discrete-time systems and difference operators, {\em
Funkts. Anal. Prilozhen,} {\bf 22}, 1-13, (1988).

\bibitem{v2}\label{v2}
A.P. Veselov, Integrable Lagrangian correspondence and the factorization of
matrix polynomials,  {\em Funkts. Anal. Prilozhen,} {\bf 25}, 38-49, (1991).

\bibitem{w1}\label{w1}
J.M. Wendlandt and J. E. Marsden, Mechanical integrators derived from a discrete
variational principle. {\em Physica D}, {\bf 106}, 223-246, (1997).

 
\end{thebibliography}
\end{document}